\documentclass[letterpaper,journal]{IEEEtran}

\usepackage{amsmath,amsfonts}
\usepackage{algorithmic}
\usepackage{algorithm}
\usepackage{array}
\usepackage[caption=false,font=normalsize,labelfont=sf,textfont=sf]{subfig}
\usepackage{textcomp}
\usepackage{stfloats}
\usepackage{url}
\usepackage{verbatim}
\usepackage{graphicx}
\usepackage{cite}

\usepackage{fancyhdr}
\fancypagestyle{arxivfirst}{%
  \fancyhf{}
  \fancyhead[C]{%
    \parbox{\textwidth}{%
      \normalfont\fontsize{8}{10}\selectfont\centering
      This work has been submitted to the IEEE for possible publication.\\
      Copyright may be transferred without notice, after which
      this version may no longer be accessible.\par
    }%
  }
  \fancyfoot[C]{\normalfont\footnotesize\thepage}

}

\graphicspath{{figures/}}

\begin{document}

\title{An Optimal Power-Flow Grouping Topology for Single-Isolation-Stage Multi-Output Solid-State Transformers}

\author{
Yi Hao,~\IEEEmembership{Student Member,~IEEE},
Krzysztof Puczko,
Levy Ferreira Costa,~\IEEEmembership{Senior Member,~IEEE},
Liangcai Shu,~\IEEEmembership{Member,~IEEE},
and Dongsheng Yang,~\IEEEmembership{Senior Member,~IEEE}%
\thanks{This work was supported by the European Union's Horizon Europe
research and innovation programme under the Marie Skłodowska-Curie
Actions (MSCA) Doctoral Networks grant agreement No. 101072414 (E2GO).}%
\thanks{Y. Hao is with the Department of Electrical Engineering,
Eindhoven University of Technology, 5612 AP Eindhoven, The Netherlands,
and also with Delta Electronics EMEA, 02-822 Warsaw, Poland
(e-mail: y.hao@tue.nl).

L. F. Costa and D. Yang are with the Department of Electrical Engineering,
Eindhoven University of Technology, 5612 AP Eindhoven, The Netherlands
(e-mail: l.costa@tue.nl; d.yang1@tue.nl).

K. Puczko is with Delta Electronics EMEA, 02-822 Warsaw, Poland
(e-mail: krzysztof.puczko@deltaww.com).

L. Shu is with Delta Electronics (Shanghai) Co., Ltd., Hangzhou, China
(e-mail: alan.shu@deltaww.com).}%
}

\markboth{}{}


\maketitle
\thispagestyle{arxivfirst}

\begin{abstract}
Electric-vehicle dc fast-charging stations and data centers require multiple high-power dc outputs with independently varying demands. Conventional multi-output solid-state transformers typically establish a common dc bus and use downstream dc/dc converters for port-level regulation and, where required, galvanic isolation. A single-isolation-stage solid-state transformer can eliminate this additional high-power interface stage by directly forming multiple isolated outputs from its submodule dc/dc stages. However, because one submodule is generally insufficient for an individual output port, multiple submodules must be connected in parallel, while inter-submodule balancing circuits redistribute power under unbalanced loading. The submodule grouping therefore determines the internal balancing-power paths. Sequential grouping can create long transfer paths and accumulated balancing power, increasing current stress and losses. This paper proposes an optimal power-flow grouping topology that interleaves submodules assigned to different output ports, thereby localizing power redistribution and reducing the power processed by the balancing circuits. Analytical models quantify the required balancing power and its reduction relative to sequential grouping. A coordinated submodule-bypass strategy is also developed, and the remaining output-power capability under different fault scenarios is evaluated. A six-submodule prototype validates the proposed topology. Under the same highly unbalanced operating conditions, the proposed topology reduces the maximum submodule dc-link voltage difference by 63\% and the peak balancing current by 14.5\%, while improving the measured efficiency by approximately 5.6\% at rated power. Stable post-fault operation is also experimentally demonstrated.
\end{abstract}


\section{Introduction}

\IEEEPARstart{S}{OLID-STATE} transformers (SSTs) have attracted considerable attention as promising interfaces between medium-voltage ac grids and low-voltage dc systems. Owing to their modular structure, high-frequency galvanic isolation, and flexible power-control capability, SSTs have been widely investigated for renewable-energy integration \cite{add1,add2,add3}, data-center power supplies \cite{add4,add5,add6}, and electric-vehicle (EV) charging stations \cite{1,2,3}. Compared with conventional line-frequency-transformer-based solutions, SSTs can also provide additional functionalities, including reactive-power compensation, harmonic mitigation, and fault protection \cite{13}. Nevertheless, system cost and conversion efficiency remain major barriers to their widespread commercialization \cite{1,4}. The design requirements of SSTs, however, vary considerably among different applications, particularly in terms of output configuration, voltage regulation, and galvanic isolation.

EV fast-charging stations represent a particularly demanding multi-output application. The voltage, current, and power demands of individual charging ports vary independently, while compliance with the relevant charging-safety requirements necessitates galvanic separation between the grid and each vehicle and, in multi-outlet stations, between simultaneously connected vehicles \cite{6,7,8}. Consequently, each charging outlet generally requires an independently regulated and galvanically isolated power interface. Data-center power systems present a different set of requirements. A common high-voltage dc bus is typically distributed to the racks, where downstream dc/dc converters provide voltage adaptation and point-of-load regulation. Nevertheless, the power demands of different racks or load groups may vary substantially with the computing workload \cite{Krein2017,datacenter2}, and distributed multi-output architectures incorporating isolated power interfaces may also be considered to enhance modularity, safety, or system reliability \cite{Chen2023,Shrestha2018}. Therefore, multi-output SSTs with independently regulated and isolated ports are especially relevant to EV fast-charging systems, while they may also provide an alternative solution for data centers depending on the adopted dc-distribution architecture.
\IEEEpubidadjcol
Among the various SST topologies, the cascaded H-bridge (CHB) structure is one of the most widely adopted configurations \cite{3}. In a conventional multi-output SST, the CHB stage is typically combined with isolated dc/dc converters whose outputs establish a common dc bus. Since this common bus cannot directly provide multiple independently regulated output ports, additional downstream dc/dc converters are required to interface with the individual loads. In EV fast-charging systems, these downstream converters must also provide the required galvanic isolation among simultaneously connected vehicles \cite{6,7,8}. 
In data-center systems, they may be isolated or nonisolated depending on the selected distribution architecture, but are still commonly required for voltage adaptation and independent load regulation \cite{Krein2017,Liang2022}. Consequently, these additional high-power conversion stages increase the component count, system cost, volume, and power losses.

\begin{figure}[!t]
\centering
\includegraphics[width=3.49in]{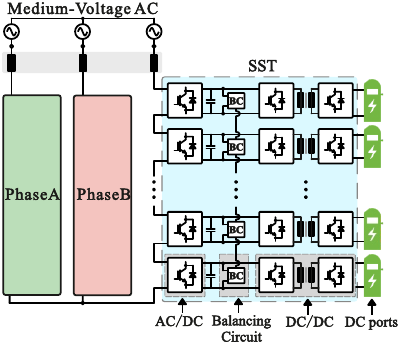}
\caption{Single-isolation-stage SST with inter-submodule balancing circuits.}
\label{fig1}
\end{figure}

To reduce the number of high-power conversion stages, several studies have investigated directly connecting loads to the dc output terminals of the SST submodules \cite{10,11}. However, because the loads connected to different submodules may demand unbalanced power, direct load connection can cause significant power imbalance among the submodules and consequently lead to severe dc-link voltage imbalance. Existing solutions to this problem can generally be categorized as control-based and hardware-based approaches. Control-based methods regulate the power processed by individual submodules through modified modulation or control strategies \cite{14,add8}. Although these methods do not require additional power hardware, their balancing capability is generally limited by the available modulation and power margins and may become insufficient under severe load imbalance. Hardware-based methods instead introduce dedicated circuits to redistribute power among the submodules. A switched-capacitor-based balancing method for dc transformers is proposed in \cite{add7}, while auxiliary circuits for achieving voltage self-balancing in modular multilevel converters are presented in \cite{15,16}. However, these methods were developed for converter structures other than CHB-based multi-output SSTs. More directly applicable to the EV fast-charging and data-center applications considered in this paper, a single-isolation-stage CHB-based SST architecture with dedicated inter-submodule balancing circuits was previously proposed in \cite{paper1,11238406}, as shown in Fig.~\ref{fig1}. In this architecture, the loads are supplied directly from the isolated dc/dc outputs of the SST submodules, eliminating the additional high-power load-side dc/dc conversion stage. The balancing circuits provide bidirectional power-transfer paths between adjacent submodules, enabling power redistribution and dc-link voltage balancing even under severe load-power imbalance. In this way, the architecture retains multiple independently regulated and galvanically isolated output ports while reducing the number of cascaded conversion and isolation stages.

However, the previous study in \cite{paper1} focused primarily on power transfer between two submodules and did not address the additional complexity introduced by a larger number of submodules in practical multi-output systems. In high-power applications, an SST typically comprises multiple submodules, several of which may be connected in parallel and operated as a group to supply the same output port. Extending the system from two submodules to a multi-submodule configuration introduces various grouping possibilities and more complex internal power-transfer paths. Consequently, the allocation of submodules among the output ports becomes an important design consideration, as it directly affects the balancing-power paths and the power processed by the balancing circuits. A straightforward sequential grouping arrangement may lead to long and unevenly distributed balancing-power paths under highly unbalanced loading conditions. As the number of submodules increases, this arrangement can result in higher balancing-current stress, increased power-processing requirements for the balancing circuits, and additional conversion losses. These unfavorable characteristics may partially offset the efficiency benefit obtained by eliminating the additional high-power load-side dc/dc conversion stage and were not investigated.

Fault-tolerant operation is another important consideration. The modular structure of CHB-based SSTs can enable continued operation after a submodule failure through redundancy and submodule bypass \cite{FAULT}. However, the introduction of inter-submodule balancing circuits creates additional power-transfer relationships among neighboring submodules. A fault in the CHB stage, balancing circuit, or isolated dc/dc stage may alter the internal power-flow paths and reduce the available output power. Therefore, it is necessary to investigate whether effective power redistribution can be maintained after submodule bypass and whether the SST can continue to operate stably under different fault conditions.

To address these issues, this paper proposes an optimal power-flow (OPF) grouping topology for single-isolation-stage SSTs with integrated balancing circuits. By optimizing the allocation of submodules among the output ports, the proposed topology shortens the internal balancing-power paths, reduces the power processed by the balancing circuits, and improves system efficiency under unbalanced loading conditions. Generalized analytical models are developed to quantify and compare the balancing-power requirements of the sequential and proposed grouping topologies. Furthermore, fault-tolerant operation after submodule bypass is investigated, and the remaining output-power capability under faults occurring in different conversion stages is analytically evaluated. A six-submodule laboratory prototype is developed to experimentally validate the proposed topology and fault-tolerant operation.

The remainder of this paper is organized as follows. Section II describes the system configuration, introduces the sequential grouping topology, and analyzes its power-flow paths. Section III presents the proposed OPF grouping topology and compares its power-flow paths with those of the sequential grouping topology. Section IV provides a quantitative analysis of the balancing-power requirements of the two grouping topologies. Section V investigates fault-tolerant operation and evaluates the remaining output-power capability under different fault conditions. Section VI presents the experimental prototype and validation results. Finally, Section VII concludes the paper.

\section{System Configuration and Submodule Grouping}
In applications such as electric-vehicle fast-charging stations and data-center power supplies, conventional three-phase SST systems commonly feed a shared dc bus, from which dedicated downstream dc/dc converters regulate and supply individual loads. In the single-isolation-stage SST architecture considered in this work, the shared dc bus and the additional load-side conversion stage are eliminated. Instead, the loads are directly supplied from the isolated dc/dc output stages of the SST submodules. Since the power capability of a single submodule is typically insufficient to meet the rating requirement of a practical output port, the outputs of multiple submodules must be connected in parallel and operated as a group to jointly supply the corresponding load. Consequently, the assignment of submodules to different output ports determines the submodule grouping configuration and, in turn, affects the internal balancing-power paths under unbalanced loading conditions. This section first introduces the straightforward sequential grouping topology and then analyzes its power-flow characteristics under unbalanced loading conditions.

\subsection{Sequential Grouping Topology}
\begin{figure}[!t]
\centering
\includegraphics[width=3.49in]{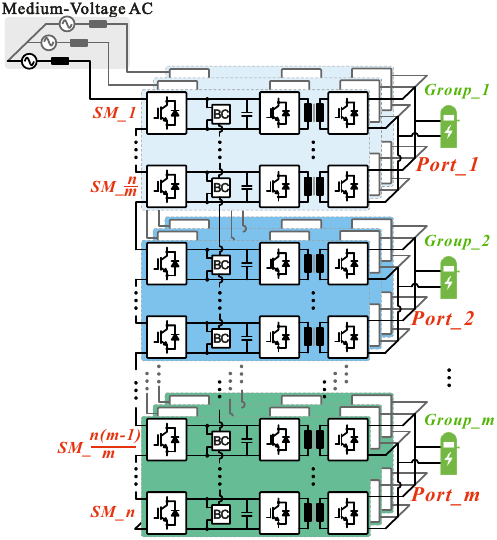}
\caption{Straightforward sequential grouping topology for single-isolation-stage SST}
\label{fig2}
\end{figure}

Based on the parallel-connection concept described above, a straightforward sequential grouping topology for multi-output SST systems is illustrated in Fig.~\ref{fig2}. Assuming that each phase of the SST consists of $n$ submodules and that the system provides $m$ output ports, where $n$ is divisible by $m$, the submodules are sequentially divided into $m$ groups, each containing $n/m$ submodules. Within each group, the LVDC terminals of the corresponding submodules are connected in parallel to form an output port. This sequential grouping topology serves as the baseline configuration considered in this paper.

This grouping topology is simple and effective, enables each output port to achieve a higher power rating through power sharing among the associated submodules. In addition, the balancing circuits enable power redistribution among submodules and groups, allowing the SST to tolerate significant load imbalance between different output ports. Consequently, the system can maintain normal operation even when some output ports are fully loaded while others remain unloaded. Such operating conditions are common in practical applications, including EV fast-charging stations and data centers, where the power demands of individual charging outlets or computing racks can vary substantially over time.

However, this operating flexibility is achieved at the expense of additional power processed by the balancing circuits. Under highly unbalanced loading conditions, the power available from the submodules associated with lightly loaded or unloaded ports must be redistributed to support the heavily loaded output ports through the balancing circuits. In the sequential grouping topology, this power may need to pass through multiple balancing-circuit stages before reaching the destination submodules. As a result, the power flow path becomes excessively long, increase the required balancing power, balancing-current stress, and associated conversion losses, thereby reducing the overall system efficiency. A more detailed analysis of the balancing power flow paths is presented in the following subsection.

\subsection{Power Flow Path Analysis}
To reveal the inherent limitations of the sequential grouping topology, its internal power-flow paths are analyzed in this subsection. For clarity, a representative system with $m=3$ output ports and $n=9$ submodules is considered. Accordingly, the nine submodules are sequentially divided into three groups, denoted as $G_1$, $G_2$, and $G_3$, with each group containing three submodules connected in parallel at their LVDC terminals.

\begin{figure}[!t]
\centering
\includegraphics[width=3.49in]{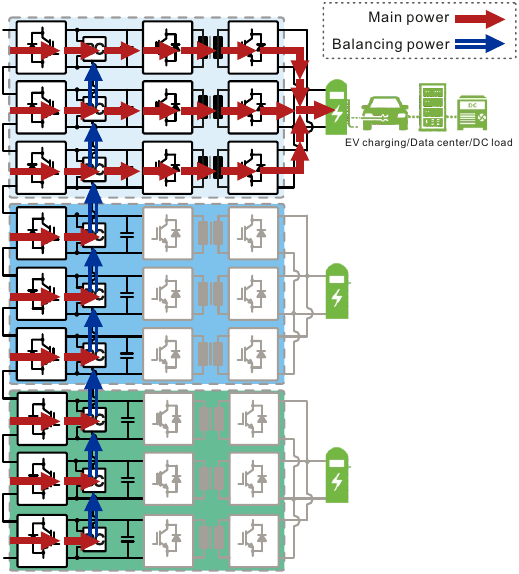}
\caption{Power flow paths in the sequential grouping topology with $n=9$ and $m=3$ ($Port_1$ loaded only)}
\label{fig4}
\end{figure}

Fig.~\ref{fig4} illustrates the power-flow paths of the sequential grouping topology when only one output port is loaded, while the other two output ports remain unloaded. Under this operating condition, the AC/DC stages of all submodules remain active to maintain the multilevel AC-side operation and share the input power. Both the AC/DC and DC/DC stages of the submodules associated with the loaded port operate normally, whereas the DC/DC stages connected to the unloaded ports are blocked. Therefore, the power processed by the AC/DC stages of the unloaded groups cannot be delivered to their local output ports and must instead be redirected to the loaded group through the balancing circuits.

Since the submodules assigned to each output port are physically arranged in consecutive positions, the balancing power is transferred between adjacent submodules in a stage-by-stage manner. The power contributed by an unloaded group located close to the loaded group passes through only a limited number of balancing-circuit stages. In contrast, the power contributed by a more distant unloaded group must traverse several intermediate submodules before reaching the destination group. Consequently, the balancing circuits located closer to the loaded group must process not only the power transferred from the adjacent group but also the accumulated power originating from more distant groups. This cumulative power-transfer characteristic results in highly uneven power stress among the balancing circuits. In particular, the balancing circuits adjacent to the loaded group may carry the combined balancing power supplied by multiple unloaded groups, while the circuits located farther away process only a portion of the transferred power. The long and concentrated power-flow paths therefore increase the peak balancing-current stress and the power processed by the balancing circuits, leading to higher conduction and switching losses. They may also aggravate the DC-link voltage differences among the submodules, thereby increasing the control effort required to maintain voltage balance.

The length of the balancing-power path and the resulting circuit stress depend strongly on the relative positions of the loaded and unloaded groups. The most severe condition generally occurs when the loaded group is located at one end of the cascaded structure, since the power available from the group at the opposite end must pass through the maximum number of balancing-circuit stages. This analysis indicates that the consecutive allocation of submodules in the sequential grouping topology inherently produces unnecessarily long and concentrated balancing-power paths under highly unbalanced loading conditions, motivating the development of a more effective submodule grouping method.

\section{Proposed Optimal Power-Flow Grouping Topology}
Based on the limitations identified for the sequential grouping topology, this section presents the proposed optimal power-flow grouping topology to shorten the balancing-power flow paths and reduce the power processed by the balancing circuits. The grouping principle and configuration are first introduced, followed by a comparative analysis of the resulting power-flow paths.
\subsection{Grouping Principle and Configuration}
\begin{figure}[!t]
\centering
\includegraphics[width=3.49in]{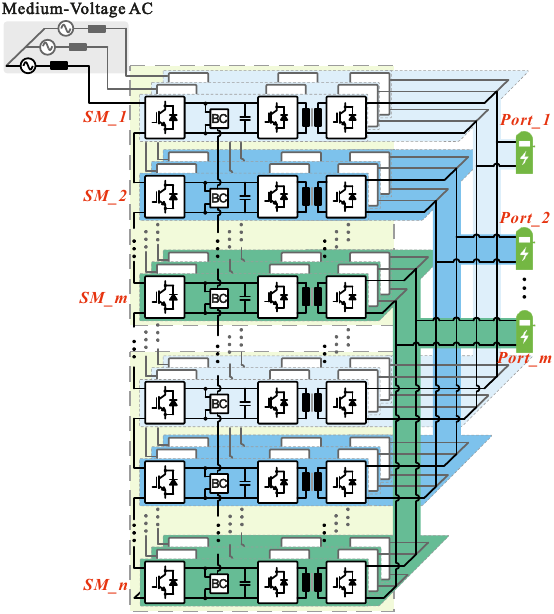}
\caption{Proposed OPF grouping topology for single-isolation-stage SST}
\label{fig3}
\end{figure}

To address the long and concentrated balancing-power paths inherent in the sequential grouping topology, an optimal power-flow grouping topology is proposed. Unlike the sequential grouping topology, in which consecutive submodules are connected in parallel to form each output port, the proposed topology adopts a cross-group interleaved parallel configuration. 
Assuming that each phase contains $n$ submodules and that the SST provides $m$ output ports, where $n$ is divisible by $m$, the submodules are first sequentially divided into $n/m$ local groups, each containing $m$ submodules. The submodules occupying the same position within different local groups are then connected in parallel at their LVDC terminals to form the corresponding output port, as illustrated in Fig.~\ref{fig3}. Consequently, the submodules assigned to each output port are distributed approximately uniformly along the cascaded structure rather than being concentrated within a continuous section.

The proposed grouping configuration retains the power-redistribution capability provided by the inter-submodule balancing circuits. Therefore, power can still be exchanged among the submodules to accommodate substantial load imbalance between the output ports. However, because the submodules associated with different output ports are interleaved throughout the SST, loaded and unloaded submodules are positioned closer to one another under highly unbalanced loading conditions.

As a result, balancing power can be exchanged primarily between neighboring submodules instead of being transferred over long distances through multiple balancing-circuit stages. The proposed arrangement also prevents the loaded and unloaded submodules from being concentrated in separate regions of the cascaded structure, thereby reducing the accumulation of balancing power in individual balancing circuits. Consequently, both the balancing-power transfer distance and the peak power processed by the balancing circuits can be reduced compared with the sequential grouping topology.

\subsection{Comparative Power-Flow Path Analysis}
\begin{figure}[!h]
\centering
\includegraphics[width=3.49in]{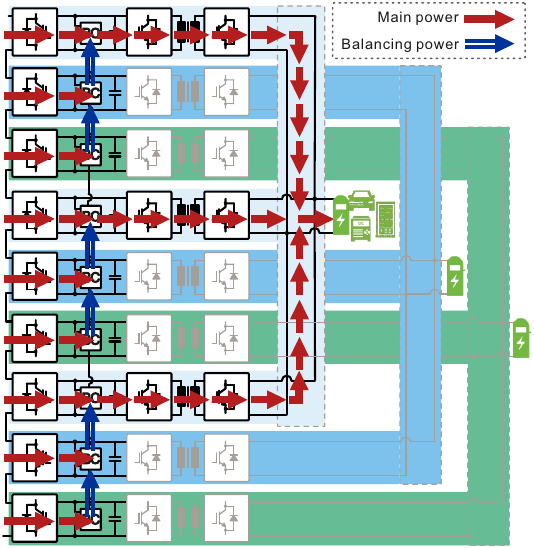}
\caption{Illustration of power flow paths in the proposed OPF grouping topology $n=9$ and $m=3$ ($Port_1$ loaded only)}
\label{fig5}
\end{figure}

For direct comparison with the sequential grouping topology analyzed in the previous section, the same representative system with $m=3$ output ports and $n=9$ submodules is considered. Fig.~\ref{fig5} illustrates the power-flow paths of the proposed OPF grouping topology when only one output port is loaded, while the other two output ports remain unloaded. As in the sequential grouping case, the DC/DC stages associated with the unloaded output ports are blocked, and the power available from the corresponding submodules must be redirected to the loaded port through the balancing circuits.

In the proposed topology, however, the submodules associated with the loaded port are interleaved with those associated with the unloaded ports. Therefore, each unloaded submodule can transfer power to a nearby loaded submodule through one or only a limited number of balancing-circuit stages. This localized redistribution mechanism substantially shortens the balancing-power paths and avoids the stage-by-stage accumulation of transferred power observed in the sequential grouping topology.

Consequently, the power processed by the most heavily stressed balancing circuits is reduced, together with the corresponding balancing-current stress and conversion losses. The proposed grouping topology is therefore expected to achieve a lower overall balancing-power requirement and improved system efficiency under highly unbalanced loading conditions. A quantitative comparison of the balancing-power requirements of the two grouping topologies is presented in the following section.

\section{Mathematical Analysis}
This section presents a quantitative analysis and comparison of the balancing-power flow paths in the sequential grouping topology and the proposed optimal power-flow grouping topology. General analytical expressions are derived to evaluate the total balancing power required for an SST with an arbitrary number of submodules and output ports.

The total balancing power required by the SST is closely related to the distance over which the balancing power must be transferred among the submodules. The operating condition in which only the first or last output port is fully loaded while all the remaining ports are unloaded represents the worst-case balancing scenario, because it results in the longest balancing-power transfer paths. Without loss of generality, the case in which only $Port\_1$ is fully loaded is considered in the following analysis.

As illustrated in Fig.~\ref{fig2}, the balancing power in the sequential grouping structure can propagate at two different levels. The first one is the power flow among different groups, where the balancing power is sequentially transferred from lightly loaded groups to the loaded group, e.g., from Group $m$ to Group $1$ in the considered worst-case condition. The second one is the power flow within the same group, where the balancing power is transferred among the submodules inside the group, e.g., from $\mathrm{SM}_{\frac{n}{m}}$ to $\mathrm{SM}_{1}$ within Group $1$.

Accordingly, the power handled by the balancing circuits can be divided into two components: the inter-group power flow $P_{\mathrm{seq\_inter}}$ and the intra-group power flow $P_{\mathrm{seq\_intra}}$. The inter-group power flow represents the total balancing power exchanged among different groups excluding the considered group itself, while the intra-group power flow represents the balancing power circulated within the considered group to maintain voltage balance among the submodules.

Assuming that each phase of the SST consists of a total of $n$ submodules and that the system contains $m$ output ports, then the number of submodules $N_g$ in one group can be expressed by (\ref{eq0}).

\begin{equation}
    \label{eq0}
    \frac{n}{m} = N g
\end{equation}

In the sequential grouping structure, the total power handled by the balancing circuits can be conveniently divided into two components: the inter-group power flow and the intra-group power flow. Assuming that each phase of the SST contains a total of $n$ submodules and that there are $m$ output ports in total, the inter-group power in the sequential grouping topology in this case is given by (\ref{eq1}), while the intra-group power is given by (\ref{eq2}), where $P_{\text{sm}}$ denotes the input power for a single submodule.

\begin{equation}
\label{eq1}
P_{\mathrm{seq\_inter}}= \sum\limits_{k = 1}^{n - N_g} {k \cdot P_{\text{sm}}}
\end{equation}

\begin{equation}
    P_{\text{seq\_intra}} = \sum_{k=1}^{N_g-1} k(m-1) P_{\text{sm}}
    \label{eq2}
\end{equation}


By simplifying (\ref{eq1}) and (\ref{eq2}), the inter-group and intra-group balancing powers can be respectively expressed as (\ref{eq3}) and (\ref{eq4}).

\begin{equation}
    P_{\text{seq\_inter}} = \frac{(n - N_g)(n - N_g + 1)}{2} P_{\text{sm}}
    \label{eq3}
\end{equation}

\begin{equation}
    P_{\text{seq\_intra}} = \frac{n(m-1)(n-m)}{2m^2} P_{\text{sm}}
    \label{eq4}
\end{equation}

From (\ref{eq0}) -  (\ref{eq4}), the total power needed to be handled by the balancing circuit in the sequential topology is given by (\ref{eq5}). 

\begin{equation}
\label{eq5}
{P_{seq}} = \frac{{{n^2}(m - 1)}}{{2m}} \cdot P_{\text{sm}}
\end{equation}

According to (\ref{eq5}), the required balancing power in the sequential grouping topology is proportional to the square of the number of submodules. Consequently, as the number of SST submodules increases, the balancing power processed by the balancing circuits grows rapidly. This quadratic growth characteristic significantly limits the scalability of the sequential grouping topology in large-scale SST systems.

By contrast, the proposed OPF grouping topology can be interpreted as a localized balancing power redistribution structure, in which balancing power transfer occurs primarily within neighboring groups. Consequently, the inter-group balancing power flow can be effectively eliminated, and the total balancing power required by the balancing circuits is significantly reduced, as expressed in (\ref{eq6}).

\begin{equation}
\label{eq6}
P_{\mathrm{OPF}} = \frac{{\rm{n}}}{m} \cdot \sum\limits_{k = 1}^{k = {\rm{m}} - 1} {k \cdot P_{\text{sm}}}  = \frac{{{\rm{n}} \cdot ({\rm{m}} - 1)}}{2} \cdot P_{\text{sm}}
\end{equation}

Assuming that the efficiency of the balancing circuits remains unchanged, the power losses are proportional to the power processed by the balancing circuits. Therefore, the required balancing power can be used as an indicator for evaluating the relative loss characteristics of different grouping topologies.

\begin{figure}[!t]
\centering
\includegraphics[width=3.49in]{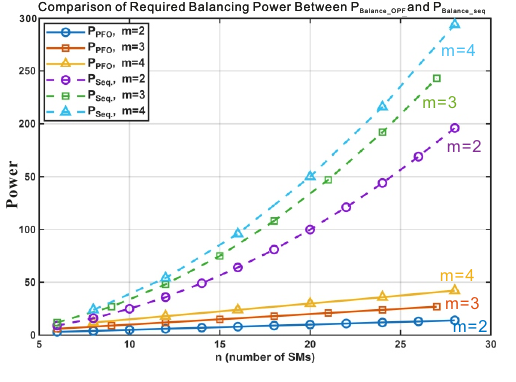}
\caption{Required normalized balancing power of the sequential and proposed OPF grouping topologies for different numbers of submodules under $m=2$, $m=3$, and $m=4$.}
\label{fig6}
\end{figure}

Fig.~\ref{fig6} compares the required balancing power of the sequential and proposed OPF grouping topologies under different numbers of submodules $n$, where the number of output ports $m$ is set to 2, 3, and 4, respectively. Only valid combinations satisfying $n>m$ and $\mathrm{mod}(n,m)=0$ are considered. As shown in the figure, for a given $m$, the required balancing power of the sequential topology increases quadratically with $n$, whereas that of the proposed OPF grouping topology increases linearly. Therefore, the difference between the two topologies becomes more pronounced as the number of submodules increases. This is because the sequential topology relies on longer balancing power propagation paths, leading to a larger cumulative balancing power. In contrast, the proposed OPF grouping topology localizes the balancing power redistribution process, thereby effectively suppressing the increase in the required balancing power.

\begin{figure}[!t]
\centering
\includegraphics[width=3.49in]{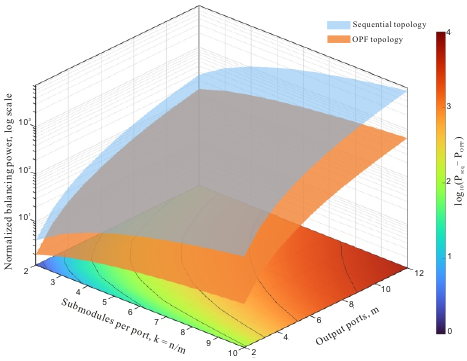}
\caption{Three-dimensional comparison of the required normalized balancing power under $n=km$. The bottom projection shows the logarithmic normalized gap.}
\label{fig6-2}
\end{figure}

To further illustrate the scalability of the proposed topology, Fig.~\ref{fig6-2} presents a three-dimensional comparison of the normalized balancing power over the valid design space. The valid combinations are represented by $n=km$, where $k=n/m$ denotes the number of submodules per output port and is an integer larger than one. The two surfaces correspond to the sequential and OPF grouping topologies, respectively, while the bottom projection shows $\log_{10}(P_{\mathrm{seq}}-P_{\mathrm{OPF}})$ to highlight the absolute gap between them. It can be observed that the proposed OPF grouping topology consistently requires lower balancing power over the considered design range. Moreover, the gap between the two topologies increases as $k$ and $m$ become larger, confirming the improved scalability of the proposed grouping topology.

\section{POWER FLOW PATH AND AVAILABILITY ANALYSIS UNDER FAULT CONDITION}
An important advantage of the proposed SST topology is that it can preserve an optimal power-flow path even under submodule fault conditions, thereby minimizing the additional balancing power and associated conduction losses after fault reconfiguration. This capability builds upon the inherent modularity and fault-tolerant nature of SSTs, which allows the overall system to continue operating when individual submodules fail. Such modularity not only enables flexible system scalability but also facilitates the implementation of redundancy. Therefore, preserving both the modular structure and the optimal power-flow characteristics is essential when introducing a new grouping topology. In this section, the fault-tolerant capability of the proposed topology is analyzed, and its performance is evaluated for submodule failures occurring at different locations.

\subsection{Modularity and Optimal Power Flow Path Preservation}
\begin{figure}[!t]
\centering
\includegraphics[width=3.49in]{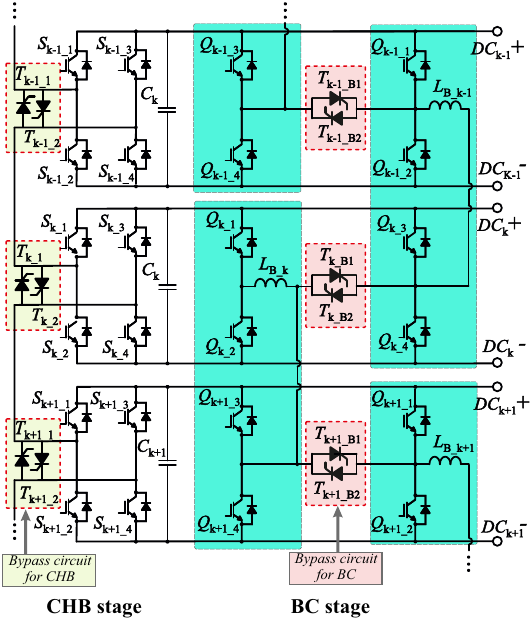}
\caption{Detailed topology of three submodules as a partial view of the proposed topology} 
\label{fig7}
\end{figure}

The detailed topology of three submodules as a partial view of the proposed SST is shown in Fig.\ref{fig7}. 
The partial view consists of the CHB stage and the balancing circuit (BC) stage. Taking the $k$-th submodule $SM_\text{k}$ as an example, switches $S_\text{k\_1}$ - $S_\text{k\_4}$ constitute the AC-side CHB circuit. Meanwhile, switches $Q_\text{k\_1}$, $Q_\text{k\_2}$, $Q_\text{k+1\_3}$, and $Q_\text{k+1\_4}$, together with the balancing inductor $L_\text{B\_k}$, form the balancing circuit that enables power transfer between adjacent submodules $SM_\text{k}$ and $SM_\text{k+1}$. Similarly, switches $Q_\text{k\_3}$ and $Q_\text{k\_4}$, together with $Q_\text{k-1\_1}$, $Q_\text{k-1\_2}$, and the balancing inductor $L_\text{B\_{k-1}}$, form the balancing circuit between $SM_\text{k-1}$ and $SM_\text{k}$.
As illustrated in Fig.\ref{fig7}, after introducing the balancing circuit, each submodule remains identical, thereby preserving the fundamental modularity of the system. This modularity is essential for scalability and enables the feasibility of redundancy-oriented designs. 

To maintain OPF and system operation under submodule fault conditions, bypass thyristors are incorporated in both the CHB stage and the balancing circuit stage. The anti-parallel thyristors $T_{k\_1}$ and $T_{k\_2}$ are connected across the AC-side CHB stage and are responsible for bypassing the CHB part when a fault occurs. Once triggered, these thyristors provide an alternative current path and isolate the faulty CHB stage from the system. This bypass mechanism is identical to that employed in conventional SST topologies.

Unlike conventional SSTs, the proposed topology additionally includes a balancing circuit between adjacent submodules. Therefore, an additional pair of anti-parallel thyristors, $T_{k\_B1}$ and $T_{k\_B2}$, is introduced to bypass the balancing circuit during fault conditions. Under normal operation, these thyristors remain blocked and do not affect power transfer. When a submodule fault is detected, they are triggered to establish a bypass path for the balancing circuit, ensuring that both the faulty submodule and its associated balancing circuit are removed from operation. As a result, continuous power delivery and reliable fault-tolerant operation can be achieved.

\begin{figure}[!t]
\centering
\subfloat[]{\includegraphics[width=2.2in]{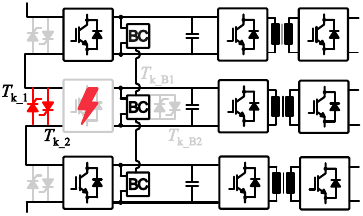}\label{fig_8a}}
\hfil
\subfloat[]{\includegraphics[width=2.2in]{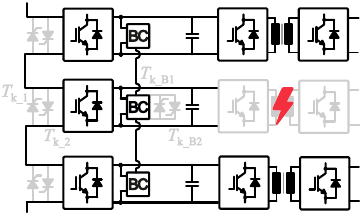}\label{fig_8b}}
\hfil
\subfloat[]{\includegraphics[width=2.2in]{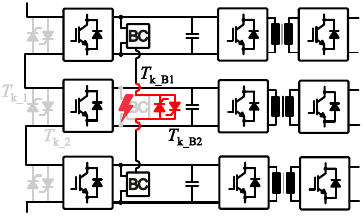}\label{fig_8c}}
\hfil
\subfloat[]{\includegraphics[width=2.2in]{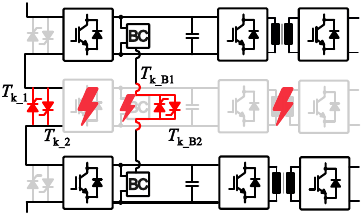}\label{fig_8d}}
\caption{Schematic diagram of three submodules as a partial view of the proposed SST:
(a) Fault happens in CHB. 
(b) Fault happens in DAB. 
(c) Fault happens in BC. 
(d) Fault occurs in the entire submodule. 
}
\label{fig8}
\end{figure}

One of the advantages of the proposed topology lies in the decoupling between the AC/DC and DC/DC conversion stages, which enables one stage to remain operational even if the other experiences a fault. This decoupling capability is primarily achieved through the integration of the balancing circuit. In conventional SSTs, any fault that occurs within a submodule typically results in bypassing the entire submodule, thereby forcing it out of operation. In the proposed topology, if a fault occurs in the CHB, the AC/DC circuit should be bypassed, but the DC/DC stage can keep operating, as shown in Fig.\ref{fig8}\,(a). This enables the SST to retain a higher output power capability under fault conditions. If a fault occurs in the DAB, the CHB circuit can still operate without being shut down, as shown in Fig.\ref{fig8}\,(b). This mitigates the fault transient current and reduces the voltage stress on the AC-side semiconductor devices. If a fault occurs in the balancing circuit, the faulty submodule can continue to operate, but the balancing circuit should be bypassed, as shown in Fig.\ref{fig8}\,(c). Fig.~\ref{fig8}(d) illustrates the fault scenario in which the entire submodule becomes faulty. In this case, both the AC-side stage and the balancing-circuit branch must be bypassed to isolate the faulty submodule from the SST system.

Overall, regardless of where a submodule fault occurs within the SST, the proposed topology preserves the optimal power-flow path. Consequently, the SST can maintain normal operation through fault bypass while minimizing the required balancing power under fault conditions.

\subsection{Fault-Tolerant Control Strategy}
\begin{figure}[!t]
\centering
\includegraphics[width=3.49in]{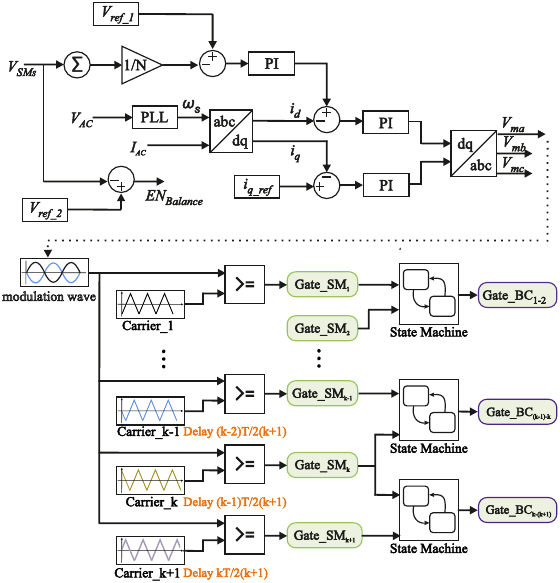}
\caption{SST control system diagram under normal operation conditions} 
\label{fig9}
\end{figure}

Under normal operating conditions, the overall control scheme of the proposed SST topology is illustrated in Fig.~\ref{fig9}. The control structure of the balancing circuit is generally consistent with that described in \cite{paper1}, which is based on a conventional double closed-loop PI control framework.

The operation of the balancing circuit depends on the monitored submodule voltages $V_{\text{SMs}}$. When the voltage of a submodule exceeds the predefined threshold $V_{\text{ref\_2}}$, the corresponding balancing function is activated to redistribute balancing power among submodules. The modulation strategy of the proposed SST is primarily based on carrier phase-shifted sinusoidal pulse-width modulation (CPS-SPWM). The triangular carriers are denoted as $Carrier\_1$ to $Carrier\_{k+1}$, where adjacent carriers are phase shifted successively by $T/2n$. Consequently, the carrier of the $(k-1)$th submodule is delayed by $(k-2)T/2n$, while the carrier of the $k$th submodule is delayed by $(k-1)T/2n$.

Similar to the method described in \cite{paper1}, the switching sequence of the balancing circuit is coordinated with the switching states of the CHB submodules, thereby establishing the required balancing power flow paths between neighboring submodules. In this manner, balancing power can be redistributed locally while maintaining the normal operation of the SST system.

As shown in Fig.~\ref{fig9}, the modulation system generates two categories of gate signals: the CHB gate signals and the balancing-circuit gate signals. The signals $Gate\_SM_1$ to $Gate\_SM_{k+1}$ are used to drive the corresponding CHB submodules, whereas the signals $Gate\_BC_{1-2}$ to $Gate\_BC_{k-(k+1)}$ are used to control the balancing circuits between adjacent submodules. Specifically, $Gate\_BC_{k-(k+1)}$ denotes the gate signal of the balancing circuit responsible for balancing power transfer between submodule $k$ and submodule $k+1$.

\begin{figure}[!t]
\centering
\includegraphics[width=3.49in]{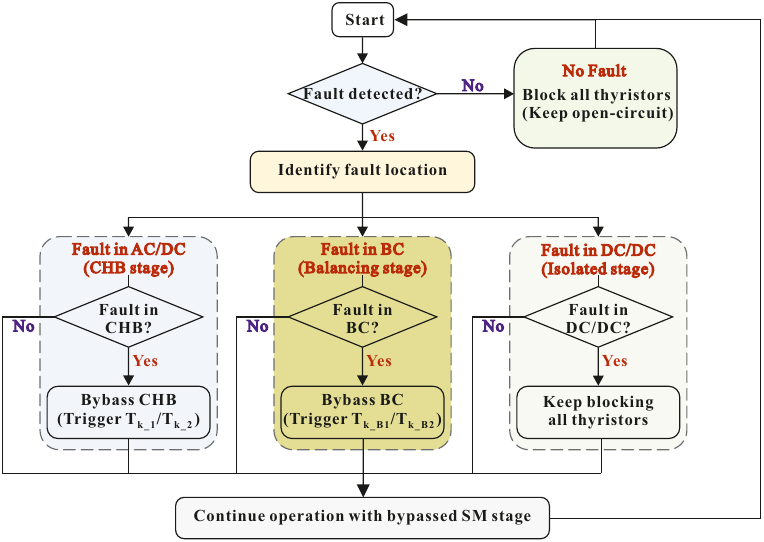}
\caption{Fault handling flowchart for the proposed topology} 
\label{fig10}
\end{figure}

Fig.~\ref{fig10} illustrates the fault detection and bypass control strategy of the proposed topology. During normal operation, all bypass thyristors remain blocked. Once a fault is detected, the fault location is first identified. If the fault occurs in the CHB stage or the balancing circuit stage, the corresponding thyristors are triggered to bypass the affected SM's conversion stage and isolate the faulty part from the system. In contrast, when a fault occurs in the isolated DC/DC stage, only the corresponding DC/DC converter is blocked, and no bypass action is required. Through this strategy, the SST can maintain operation in fault conditions.

\begin{figure}[!t]
\centering
\includegraphics[width=3.49in]{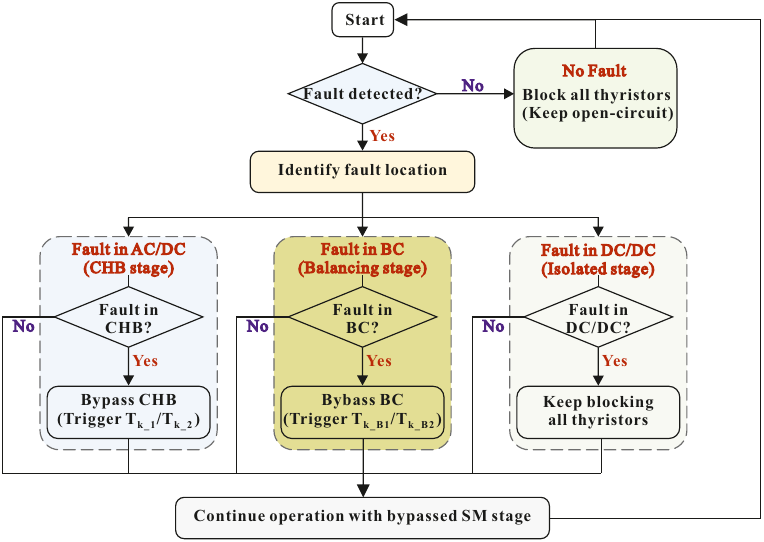}
\caption{SST control system diagram under fault operation conditions} 
\label{fig11}
\end{figure}

After the bypass function is activated, to maintain the balancing power transfer capability after submodule faults, both the CPS-SPWM carrier allocation and the balancing-circuit gate signals must be reconfigured according to the remaining active submodules.
Fig.~\ref{fig11} illustrates the control scheme under fault operating conditions, where the entire $SM_k$ submodule is bypassed after a fault, as illustrated in Fig.~\ref{fig8}(d). The corresponding gate signal $Gate\_SM_k$ is blocked to isolate the faulty submodule from the SST system. Meanwhile, the carrier signal $Carrier_k$, which was originally assigned to the faulty submodule, is reassigned to generate the switching signals of the remaining active submodules. Consequently, the CPS-SPWM phase-shift relationship among the carriers is reconfigured according to the updated number of operating submodules. Taking the $k$th submodule as an example, the carrier phase-shift angle is adjusted from $(k-1)T/2(k+1)$ to $(k-1)T/2k$ after fault reconfiguration.

For the balancing-circuit gate signals, the original signal $Gate\_BC_{k-(k+1)}$, which is responsible for balancing power transfer between the $k$th and $(k+1)$th submodules, is also blocked after the $k$th submodule is bypassed. Consequently, a new balancing-circuit gate signal, denoted as $Gate\_BC_{(k-1)-(k+1)}$ is reconstructed to establish the balancing power flow path between the $(k-1)$th and $(k+1)$th submodules. In this manner, the voltage balancing function can be maintained among the remaining active submodules after fault reconfiguration.

As a result, the proposed fault reconfiguration strategy enables the SST system to preserve localized balancing power transfer capability and maintain higher output power capability under fault conditions.

\subsection{Impact of Fault Locations on Output Power Capability}
Due to the introduction of the balancing circuit and the OPF grouping topology, the output power capability for the proposed topology under fault conditions differs from that for conventional SSTs. Since faults occurring at different locations exert distinct impacts on the overall performance of the SST, it is necessary to analyze each fault scenario separately.
The SST can be divided into three major parts: the cascaded H-bridge, the balancing circuit, and the dual active bridge. 
Moreover, due to the concept of grouping, fault analysis should further distinguish whether the fault occurs within the load-connected group or within the no-load group.
The following analysis is carried out using the most extreme case as an example, when only the first output port is connected to a load, while all other ports remain unloaded.

Benefiting from the aforementioned redundancy capability, faults occurring in the AC/DC stage or balancing circuit of an unloaded submodule only require the corresponding faulty submodule to be bypassed and isolated. Consequently, the overall SST system only loses the power contribution of a single submodule. Moreover, faults in the DC/DC stage of unloaded submodules do not affect the output power capability of the SST system.
Faults occurring in loaded submodules are generally more severe. If a fault occurs in the AC/DC stage of a loaded submodule, only the power contribution of the corresponding submodule is lost. However, if a fault occurs in the balancing circuit, the remaining submodules within the same group can no longer transfer balancing power to the faulty submodule. As a result, the power contribution of the entire remaining group is lost. The most severe case occurs when a fault takes place in the DC/DC stage of a loaded submodule, since the output power capability of the entire group will then be completely lost.

\begin{figure}[!h]
\centering
\subfloat[]{\includegraphics[width=3.4in]{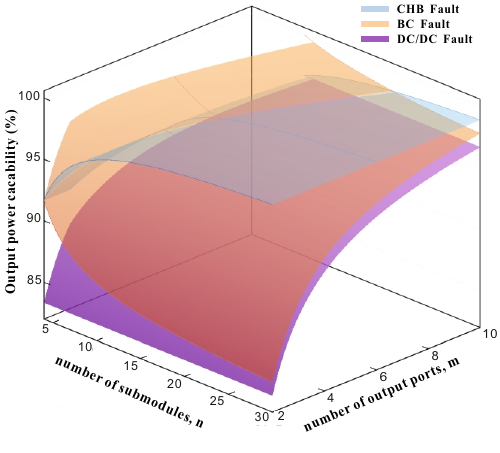}\label{fig table_a}}
\hfil
\subfloat[]{\includegraphics[width=3.4in]{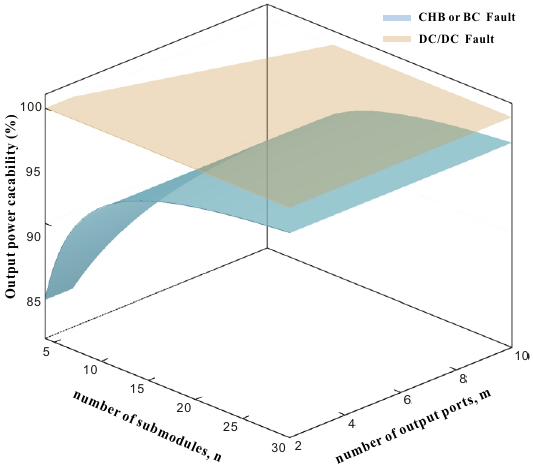}\label{fig table_b}}
\hfil
\caption{Post-fault power capability for faults occurring in the CHB, balancing circuit , and DC/DC stage of (a) loaded submodules and (b) unloaded submodules.}

\label{fig table}
\end{figure}

The post-fault output-power capability is illustrated in Fig.~\ref{fig table}, where $n$ denotes the number of submodules per phase, resulting in a total of $3n$ submodules in the three-phase SST, and $m$ denotes the number of output ports. Fig.~\ref{fig table_a} presents the output-power capability for faults occurring at different circuit locations within a loaded submodule, whereas Fig.~\ref{fig table_b} presents the corresponding results for faults occurring within an unloaded submodule. It can be observed that if a fault occurs in the dc/dc circuit of a no-load group, it has no impact on the output power. In contrast, a fault in the dc/dc stage of a load-carrying submodule results in the most severe reduction in system performance. For faults occurring at the other considered locations, the corresponding post-fault output-power capability of the SST can also be identified from the figure. Overall, the proposed topology retains a relatively high output-power capability under the investigated fault conditions.

\section{Experimental Validation}
\begin{figure}[!t]
\centering
\includegraphics[width=3.49in]{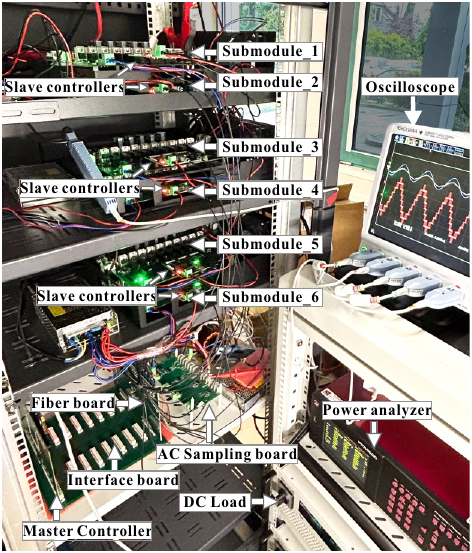}
\caption{Photograph of the six-module SST prototype.} 
\label{fig12}
\end{figure}

To verify the proposed topology presented in this paper as shown in Fig.~\ref{fig3}, an SST prototype consisting of six submodules was developed, where $n=6$ and $m=3$. Each submodule contains a full-bridge circuit comprising the CHB stage, a balancing circuit for balancing power transfer, and dedicated bypass switching devices for fault isolation and reconfiguration. The control system of the prototype is composed of a master controller based on the ZYNQ-7000 platform, fiber-optic communication boards, and slave controllers implemented using Spartan\texttrademark~7 FPGAs. A photograph of the experimental prototype is shown in Fig.~\ref{fig12}, and the parameters are given in Table \ref{Table1}.

\begin{table}[!b]
\centering
\renewcommand{\arraystretch}{1.5} 
\caption{Prototype Parameters}
\label{Table1}
\begin{tabular}{lll}
\hline
Parameter                     & Symbol & Value \\ \hline
AC Voltage                    & $U_\text{AC}$             & 250 V  \\
DC Voltage                    & $U_\text{DC}$             & 50 V   \\
Switching frequency           & $f_\text{CHB}$            & 5 kHz    \\
Inductance of bridge inductor & $L_\text{AC}$             & 5 mH    \\
Submodule DC capacitor        & $C_\text{DC}$             & $3000\ \mu\text{F}$  \\
Balancing inductor            & $L_\text{B}$              & $10\ \mu\text{H}$   \\
Rated Power for single SM     & $P_\text{R}$              & 250 W    \\ \hline
\end{tabular}
\end{table}

\begin{figure}[!t]
\centering
\includegraphics[width=3.49in]{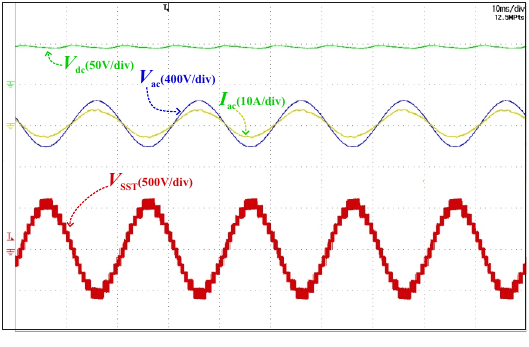}
\caption{Steady-state waveforms of the SST: $V_{\mathrm{SST}}$, $V_{\mathrm{AC}}$, $I_{\mathrm{AC}}$, and $V_{\mathrm{dc}}$.}
\label{fig13}
\end{figure}

The steady-state operation capability of the proposed topology under unbalanced loading conditions was first verified. In the experimental setup, six submodules were configured to form three output ports, where only the first output port was loaded. Specifically, only submodule 1 and submodule 4 delivered output power, while the remaining submodules operated under unloaded conditions. The corresponding steady-state waveforms are shown in Fig.~\ref{fig13}, including the input supply voltage $V_{ac}$, the SST port voltage $V_{SST}$, the AC-side input current $I_{ac}$ and the output voltage $V_{DC}$. The steady-state waveforms demonstrate that, even under extremely unbalanced loading conditions, the SST with the OPF grouping topology operates stably while retaining a 13-level AC-side voltage waveform.

\begin{figure}[!t]
\centering
\includegraphics[width=3.49in]{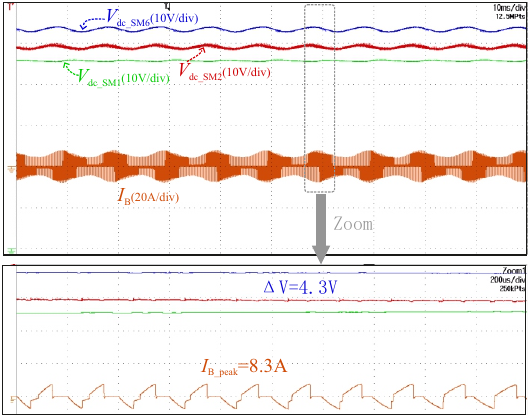}
\caption{Experimental waveforms under the sequential grouping topology: $V_{\mathrm{dc},\mathrm{SM1}}$, $V_{\mathrm{dc},\mathrm{SM2}}$, $V_{\mathrm{dc},\mathrm{SM6}}$, and $I_{\mathrm{B}}$.}
\label{fig14}
\end{figure}

\begin{figure}[!t]
\centering
\includegraphics[width=3.49in]{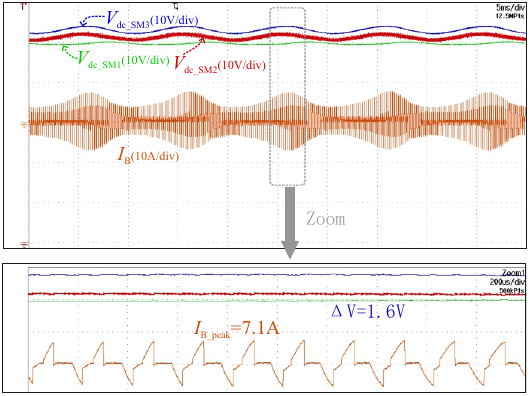}
\caption{Experimental waveforms under the OPF grouping topology: $V_{\mathrm{dc},\mathrm{SM1}}$, $V_{\mathrm{dc},\mathrm{SM2}}$, $V_{\mathrm{dc},\mathrm{SM3}}$, and $I_{\mathrm{B}}$.}
\label{fig15}
\end{figure}

For comparison, Fig.~\ref{fig14} and Fig.~\ref{fig15} show the submodule DC-link voltages and balancing-circuit currents obtained with the sequential grouping and OPF grouping topologies, respectively. Under the sequential grouping topology, Fig.~\ref{fig14} shows $V_{\mathrm{dc},\mathrm{SM1}}$, $V_{\mathrm{dc},\mathrm{SM2}}$, and $V_{\mathrm{dc},\mathrm{SM6}}$, together with the balancing current $I_{\mathrm{B}}$ associated with $\mathrm{SM2}$. The maximum voltage difference among the submodules reaches $4.3~\mathrm{V}$, while the peak balancing current reaches $8.3~\mathrm{A}$, indicating a relatively high balancing effort under this operating condition.

In contrast, Fig.~\ref{fig15} shows $V_{\mathrm{dc},\mathrm{SM1}}$, $V_{\mathrm{dc},\mathrm{SM2}}$, and $V_{\mathrm{dc},\mathrm{SM3}}$, together with the balancing-circuit current $I_{\mathrm{B}}$ associated with $\mathrm{SM1}$ under the OPF grouping topology. Under the same load condition, the maximum voltage difference between the submodules is reduced from $0.086~\mathrm{p.u.}$ to $0.032~\mathrm{p.u.}$, corresponding to a reduction of approximately $63\%$. In addition, the peak balancing current decreases from $8.3~\mathrm{A}$ to $7.1~\mathrm{A}$, representing a reduction of approximately $14.5\%$. This improvement is mainly attributed to the optimized balancing-power flow paths provided by the OPF grouping topology, which reduce the amount of power that must be transferred through the balancing circuits.

\begin{figure}[!t]
\centering
\includegraphics[width=3.49in]{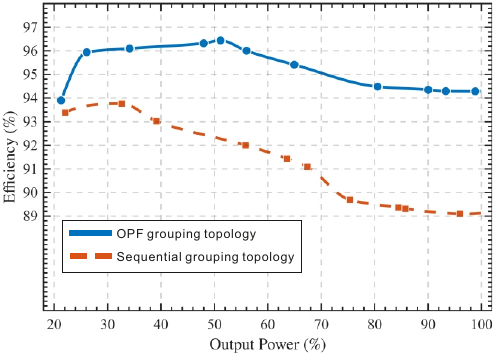}
\caption{Efficiency comparison between the proposed OPF grouping topology and the sequential grouping topology}
\label{fig16}
\end{figure}

Fig.~\ref{fig16} compares the measured efficiency curves of the proposed OPF grouping topology and the sequential grouping topology. The proposed OPF grouping topology achieves higher overall efficiency over the measured operating range. This improvement is consistent with the reduced balancing-circuit current stress and power-processing requirement demonstrated above, which result in lower losses in the balancing circuits.

\begin{figure}[!t]
\centering
\includegraphics[width=3.49in]{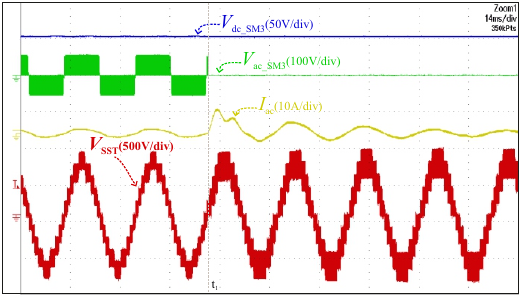}
\caption{Voltage and current during bypass: $V_{\mathrm{SST}}$, $V_{\mathrm{AC},\mathrm{SM3}}$, $V_{\mathrm{dc},\mathrm{SM3}}$, and $I_{\mathrm{AC}}$}
\label{fig17}
\end{figure}

Fig.~\ref{fig17} shows the experimental waveforms during the bypass operation, including $V_{\mathrm{SST}}$, $V_{\mathrm{AC},\mathrm{SM3}}$, $V_{\mathrm{dc},\mathrm{SM3}}$, and $I_{\mathrm{AC}}$. At $t_1$, the faulty submodule $SM3$ is bypassed. The results demonstrate that the OPF grouping topology can keep the balancing power flow path during the fault, thereby enabling the SST to maintain stable operation.

\section{Conclusion}
In this paper, an optimal power-flow grouping topology is proposed for solid-state transformers employing a single isolation stage and inter-submodule balancing circuits. By optimizing the power flow paths among the submodules, the proposed topology reduces the balancing power that must be processed by the balancing circuits under unbalanced load conditions. Compared with the straightforward sequential grouping topology, the reduction in the required balancing power is quantitatively derived, and its relationship with the number of submodules and output ports is analytically established. A corresponding control strategy is also developed to coordinate the power conversion and inter-submodule power transfer.
A six-submodule SST prototype was developed to experimentally validate the proposed topology. Under the same highly unbalanced operating condition, the OPF grouping topology reduced the maximum submodule DC-link voltage difference by 63\% and the peak balancing current by 14.5\%. The reduced balancing-power requirement also resulted in higher measured system efficiency over the investigated operating range, with an efficiency improvement of approximately 5.6\% at rated power. Moreover, the proposed topology retained the full multilevel AC-terminal voltage waveform under extremely unbalanced loading conditions. Following a submodule fault and bypass operation, the OPF grouping topology enabled the SST to continue operating stably.
These results verify that the proposed topology can simultaneously improve power-flow distribution, voltage balancing, and operating efficiency in single-isolation-stage multiport SSTs.

\bibliographystyle{IEEEtran} 
\bibliography{ref}

@ARTICLE{paper1,
  author={Hao, Yi and Costa, Levy Ferreira and Shu, Liangcai and Puczko, Krzysztof and Yang, Dongsheng},
  journal={IEEE Transactions on Power Electronics}, 
  title={A Novel Solid-State Transformer Structure With a Single Galvanic Isolation Stage for Multioutlet EV Fast Charging Stations}, 
  year={2026},
  volume={41},
  number={4},
  pages={6733-6744},
  doi={10.1109/TPEL.2025.3617768}}

@INPROCEEDINGS{11238406,
  author={Hao, Yi and Costa, Levy Ferreira and Shu, Liangcai and Puczko, Krzysztof and Yang, Dongsheng},
  booktitle={2025 Energy Conversion Congress \& Expo Europe (ECCE Europe)}, 
  title={A Single-Stage Solid-State Transformers with Balancing Circuit for Electrical Vehicle Fast Charging Stations}, 
  year={2025},
  volume={},
  number={},
  pages={1-6},
  doi={10.1109/ECCE-Europe62795.2025.11238406}}

@ARTICLE{1,
  author={Huber, Jonas E. and Kolar, Johann W.},
  journal={IEEE Transactions on Smart Grid}, 
  title={Applicability of Solid-State Transformers in Today’s and Future Distribution Grids}, 
  year={2019},
  volume={10},
  number={1},
  pages={317-326},
  doi={10.1109/TSG.2017.2738610}}

@ARTICLE{2,
  author={She, Xu and Huang, Alex Q. and Burgos, Rolando},
  journal={IEEE Journal of Emerging and Selected Topics in Power Electronics}, 
  title={Review of Solid-State Transformer Technologies and Their Application in Power Distribution Systems}, 
  year={2013},
  volume={1},
  number={3},
  pages={186-198},
  doi={10.1109/JESTPE.2013.2277917}}

@ARTICLE{3,
  author={Ruiz, Felipe and Perez, Marcelo A. and Espinosa, José R. and Gajowik, Tomasz and Stynski, Sebastian and Malinowski, Mariusz},
  journal={IEEE Industrial Electronics Magazine}, 
  title={Surveying Solid-State Transformer Structures and Controls: Providing Highly Efficient and Controllable Power Flow in Distribution Grids}, 
  year={2020},
  volume={14},
  number={1},
  pages={56-70},
  doi={10.1109/MIE.2019.2950436}}

@INPROCEEDINGS{4,
  author={Huber, Jonas E. and Kolar, Johann W.},
  booktitle={2014 IEEE Energy Conversion Congress and Exposition (ECCE)}, 
  title={Volume/weight/cost comparison of a 1MVA 10 kV/400 V solid-state against a conventional low-frequency distribution transformer}, 
  year={2014},
  volume={},
  number={},
  pages={4545-4552},
  doi={10.1109/ECCE.2014.6954023}}

@misc{6,
  author       = {International Electrotechnical Commission},
  title        = {{IEC 61851-23:2023: Electric vehicle conductive charging system – Part 23: DC electric vehicle charging station}},
  year         = {2023},
  howpublished = {International Standard, IEC},
  note         = {Available from: https://webstore.iec.ch/publication/69988},
}

@misc{7,
  author       = {International Electrotechnical Commission},
  title        = {{IEC 61851-1:2017: Electric vehicle conductive charging system – Part 1: General requirements}},
  year         = {2017},
  howpublished = {International Standard, IEC},
  note         = {Available from: https://webstore.iec.ch/publication/26196},
}

@misc{8,
  author       = {Standardization Administration of China},
  title        = {{GB/T 18487.5—2024: Electric vehicle conductive charging system—Part 5: Multi-vehicle DC charging requirements}},
  year         = {2024},
  howpublished = {Chinese National Standard},
  note         = {In Chinese},
}

@ARTICLE{10,
  author={Vasiladiotis, Michail and Rufer, Alfred},
  journal={IEEE Transactions on Industrial Electronics}, 
  title={A Modular Multiport Power Electronic Transformer With Integrated Split Battery Energy Storage for Versatile Ultrafast EV Charging Stations}, 
  year={2015},
  volume={62},
  number={5},
  pages={3213-3222},
  doi={10.1109/TIE.2014.2367237}}

@INPROCEEDINGS{11,
  author={De Seram, Ruvini and Lulbadda, Kushan Tharuka and Sidhu, Tarlochan and Williamson, Sheldon S.},
  booktitle={2024 4th International Conference on Smart Grid and Renewable Energy (SGRE)}, 
  title={A Solid State Transformer-based Charging Station for Catering Heavy/Medium and Light Electric Vehicles}, 
  year={2024},
  volume={},
  number={},
  pages={1-6},
  doi={10.1109/SGRE59715.2024.10428728}}

@ARTICLE{13,
  author={Ahmad, Adnan and Qin, Zian and Wijekoon, Thiwanka and Bauer, Pavol},
  journal={IEEE Open Journal of the Industrial Electronics Society}, 
  title={An Overview on Medium Voltage Grid Integration of Ultra-Fast Charging Stations: Current Status and Future Trends}, 
  year={2022},
  volume={3},
  number={},
  pages={420-447},
  doi={10.1109/OJIES.2022.3179743}}

@ARTICLE{14,
  author={Zhao, Tiefu and Wang, Gangyao and Bhattacharya, Subharshish and Huang, Alex Q.},
  journal={IEEE Transactions on Power Electronics}, 
  title={Voltage and Power Balance Control for a Cascaded H-Bridge Converter-Based Solid-State Transformer}, 
  year={2013},
  volume={28},
  number={4},
  pages={1523-1532},
  doi={10.1109/TPEL.2012.2216549}}

@ARTICLE{15,
  author={Gao, Congzhe and Lv, Jingliang},
  journal={IEEE Transactions on Power Delivery}, 
  title={A New Parallel-Connected Diode-Clamped Modular Multilevel Converter With Voltage Self-Balancing}, 
  year={2017},
  volume={32},
  number={3},
  pages={1616-1625},
  doi={10.1109/TPWRD.2017.2670662}}

@ARTICLE{16,
  author={Shu, Hongchun and Lei, Shunguang and Tian, Xincui},
  journal={IEEE Access}, 
  title={A New Topology of Modular Multilevel Converter With Voltage Self-Balancing Ability}, 
  year={2019},
  volume={7},
  number={},
  pages={184786-184796},
  doi={10.1109/ACCESS.2019.2958857}}

@ARTICLE{add1,
  author={Liu, Tao and Yang, Xu and Chen, Wenjie and Li, Yang and Xuan, Yang and Huang, Lang and Hao, Xiang},
  journal={IEEE Transactions on Power Electronics}, 
  title={Design and Implementation of High Efficiency Control Scheme of Dual Active Bridge Based 10 kV/1 MW Solid State Transformer for PV Application}, 
  year={2019},
  volume={34},
  number={5},
  pages={4223-4238},
  doi={10.1109/TPEL.2018.2864657}}

@ARTICLE{add2,
  author={Zheng, Liran and Kandula, Rajendra Prasad and Divan, Deepak},
  journal={IEEE Transactions on Power Electronics}, 
  title={Multiport Control With Partial Power Processing in Solid-State Transformer for PV, Storage, and Fast-Charging Electric Vehicle Integration}, 
  year={2023},
  volume={38},
  number={2},
  pages={2606-2616},
  doi={10.1109/TPEL.2022.3211000}}

@ARTICLE{add3,
  author={Zhuang, Yizhan and Liu, Fei and Huang, Yanhui and Wang, Shiwen and Pan, Shangzhi and Zha, Xiaoming and Diao, Xiaoguang},
  journal={IEEE Transactions on Power Electronics}, 
  title={A Multiport DC Solid-State Transformer for MVDC Integration Interface of Multiple Distributed Energy Sources and DC Loads in Distribution Network}, 
  year={2022},
  volume={37},
  number={2},
  pages={2283-2296},
  doi={10.1109/TPEL.2021.3105528}}

@ARTICLE{add4,
  author={Chen, Yenan and Shi, Keyan and Chen, Min and Xu, Dehong},
  journal={IEEE Journal of Emerging and Selected Topics in Power Electronics}, 
  title={Data Center Power Supply Systems: From Grid Edge to Point-of-Load}, 
  year={2023},
  volume={11},
  number={3},
  pages={2441-2456},
  doi={10.1109/JESTPE.2022.3229063}}

@ARTICLE{add5,
  author={Rothmund, Daniel and Guillod, Thomas and Bortis, Dominik and Kolar, Johann W.},
  journal={IEEE Journal of Emerging and Selected Topics in Power Electronics}, 
  title={99\% Efficient 10 kV SiC-Based 7 kV/400 V DC Transformer for Future Data Centers}, 
  year={2019},
  volume={7},
  number={2},
  pages={753-767},
  doi={10.1109/JESTPE.2018.2886139}}

@ARTICLE{add6,
  author={Zhu, Lin and Li, Chushan and Yang, Huan and Li, Wuhua and He, Xiangning},
  journal={IEEE Transactions on Power Electronics}, 
  title={Enhanced Fault-Tolerant Operation Strategy for Cascaded Vienna Front-End Based Solid-State Transformer}, 
  year={2023},
  volume={38},
  number={10},
  pages={12139-12152},
  doi={10.1109/TPEL.2023.3294186}}

@ARTICLE{add7,
  author={Guan, Renfeng and He, Zhixing and Xiao, Muxuan and Liang, Ruxin and Xiao, Yu and Li, Zongjian and Wang, Lei and Luo, An},
  journal={IEEE Transactions on Power Electronics}, 
  title={A Medium Voltage Input Multiport Isolated Output DC Transformer With Power Self-Balancing and Output Fault Isolation}, 
  year={2023},
  volume={38},
  number={4},
  pages={4771-4786},
  doi={10.1109/TPEL.2022.3230850}}

@ARTICLE{add8,
  author={Wang, Liang and Zhang, Donglai and Wang, Yi and Wu, Bin and Athab, Hussain S.},
  journal={IEEE Transactions on Power Electronics}, 
  title={Power and Voltage Balance Control of a Novel Three-Phase Solid-State Transformer Using Multilevel Cascaded H-Bridge Inverters for Microgrid Applications}, 
  year={2016},
  volume={31},
  number={4},
  pages={3289-3301},
  doi={10.1109/TPEL.2015.2450756}}

@ARTICLE{datacenter2,
  author={Dayarathna, Miyuru and Wen, Yonggang and Fan, Rui},
  journal={IEEE Communications Surveys \& Tutorials}, 
  title={Data Center Energy Consumption Modeling: A Survey}, 
  year={2016},
  volume={18},
  number={1},
  pages={732-794},
  doi={10.1109/COMST.2015.2481183}}

@ARTICLE{Chen2023,
  author={Chen, Shuohang and Zhang, Guidong and Yu, Samson S. and Mei, Yi and Zhang, Yun},
  journal={Chinese Journal of Electrical Engineering}, 
  title={A Review of Isolated Bidirectional DC-DC Converters for Data Centers}, 
  year={2023},
  volume={9},
  number={4},
  pages={1-22},
  doi={10.23919/CJEE.2023.000034}}

@ARTICLE{Shrestha2018,
  author={Shrestha, Bijen R. and Tamrakar, Ujjwol and Hansen, Timothy M. and Bhattarai, Bishnu P. and James, Sean and Tonkoski, Reinaldo},
  journal={IEEE Access}, 
  title={Efficiency and Reliability Analyses of AC and 380 V DC Distribution in Data Centers}, 
  year={2018},
  volume={6},
  number={},
  pages={63305-63315},
  doi={10.1109/ACCESS.2018.2877354}}

@ARTICLE{Krein2017,
  author={Krein, Philip T.},
  journal={CPSS Transactions on Power Electronics and Applications}, 
  title={Data center challenges and their power electronics}, 
  year={2017},
  volume={2},
  number={1},
  pages={39-46},
  doi={10.24295/CPSSTPEA.2017.00005}}

@ARTICLE{Liang2022,
  author={Liang, Jiawei and Wang, Liang and Fu, Minfan and Liang, Junrui and Wang, Haoyu},
  journal={CPSS Transactions on Power Electronics and Applications}, 
  title={Overview of Voltage Regulator Modules in 48 V Bus-Based Data Center Power Systems}, 
  year={2022},
  volume={7},
  number={3},
  pages={283-299},
  doi={10.24295/CPSSTPEA.2022.00026}}

@ARTICLE{FAULT,
  author={Tian, Jie and Mao, Chengxiong and Wang, Dan and Nie, Shaoxiong and Yang, Yun},
  journal={IEEE Transactions on Power Electronics}, 
  title={A Short-Time Transition and Cost Saving Redundancy Scheme for Medium-Voltage Three-Phase Cascaded H-Bridge Electronic Power Transformer}, 
  year={2018},
  volume={33},
  number={11},
  pages={9242-9252},
  doi={10.1109/TPEL.2018.2794558}}

\vfill

\end{document}